\documentclass[a4paper]{jpconf}
\usepackage{graphicx}
\usepackage{mathtools}
\usepackage{hyperref}
\usepackage{color}
\begin{document}
\title{Understanding two same-sign and three leptons with $b$-jets in four top quark events at the LHC}

\author{Thuso Mathaha$^1$, Abhaya Kumar Swain$^1$, Mukesh Kumar$^1$,\\ Xifeng Ruan$^1$, Bruce Mellado$^{1,2}$}

\address{$^1$School of Physics and Institute for Collider Particle Physics, University of the
Witwatersrand, Johannesburg, Wits 2050, South Africa}
\address{$^2$iThemba LABS, National Research Foundation, PO Box 722, Somerset West 7129,
South Africa}

\ead{thuso.stephen.mathata@cern.ch}

\begin{abstract}
The top quark is the heaviest known elementary particle of the Standard Model (SM) of particle physics and, therefore, it is expected to have large couplings to hypothetical new physics in many models beyond the SM (BSM). 
Various studies have predicted the presence of multi-lepton anomalies at the LHC. One of those anomalies is the excess production of two same-sign leptons and three isolated leptons in association with $b$-jets. These are reasonably well described by a 2HDM+$S$ model, where $S$ is a singlet scalar. Both the ATLAS and CMS experiments have reported sustained excesses in these final states. This includes corners of the phase-space where production of top quark pairs in association with a $W$ boson contributes to. Here, we investigate the production of two same-sign and three leptons from the production of four top quark final states.
Our focus is on understanding the differences between the SM and BSM production mechanisms of four top quarks from $t\overline{t} A$ ($A \rightarrow t\overline{t}$) using Machine Leaning techniques with twelve discriminating kinematic variables.
\end{abstract}

\section{Introduction}
The exploration of a Higgs boson ($h$) at the Large Hadron Collider (LHC) by A Toroidal LHC ApparatuS (ATLAS)~\cite{ATLAS:2012yve} and Compact Muon Solenoid (CMS)~\cite{CMS:2012qbp} has opened a new window of opportunity for the community of particle physics. The measurement properties relating to the Higgs boson have illustrated the compatibility with those predicted by the Standard Model (SM)~\cite{CMS:2012vby,ATLAS:2013xga}. Be that as it may, the possibility of the existence of additional scalar bosons is not excluded provided that their mixing with the SM Higgs boson is adequately small. Subsequently with the higher luminosity, the focus has shifted towards understanding the couplings of the Higgs boson to the SM particles and searching for new particles. Thus, leaving no stone unturned to search for new particles/interactions at the LHC. In doing so the LHC has already reported a few prelusive hints/anomalies in its current data which needs immediate attention. 
In this context a scalar singlet $S$ was introduced in conjunction with a 2HDM model in Ref.~\cite{vonBuddenbrock:2015ema,vonBuddenbrock:2016rmr} to explain some features of the Run 1 LHC data, referred to as the 2HDM+$S$ model. The model predicts the emergence of multi-lepton anomalies that have been verified in Refs.~\cite{vonBuddenbrock:2017gvy,Buddenbrock:2019tua,vonBuddenbrock:2020ter,Hernandez:2019geu}, where a possible candidate of $S$ has been reported in Ref.~\cite{Crivellin:2021ubm}. The model can further elaborate on multiple anomalies in astro-physics if it is complemented by a candidate of a Dark matter~\cite{Beck:2021xsv}. It can be easily extended~\cite{Sabatta:2019nfg} to account for the $4.2\sigma$ anomaly $g-2$ of the muon~\cite{Muong-2:2021ojo,Aoyama:2020ynm}.

\begin{figure}[t]
\centering
\includegraphics[width=.22\textwidth,height=0.18\textwidth]{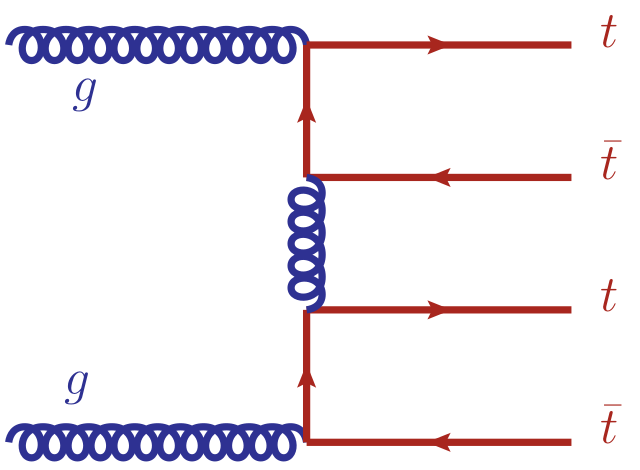}
\includegraphics[width=.22\textwidth,height=0.18\textwidth]{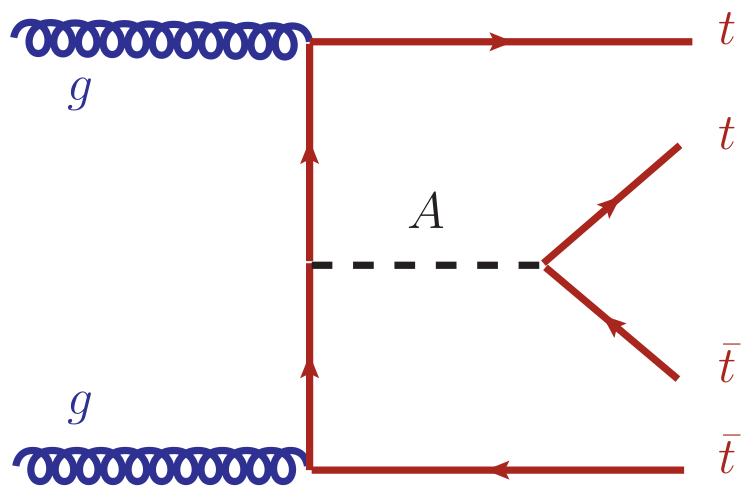}
\includegraphics[width=.25\textwidth,height=0.18\textwidth]{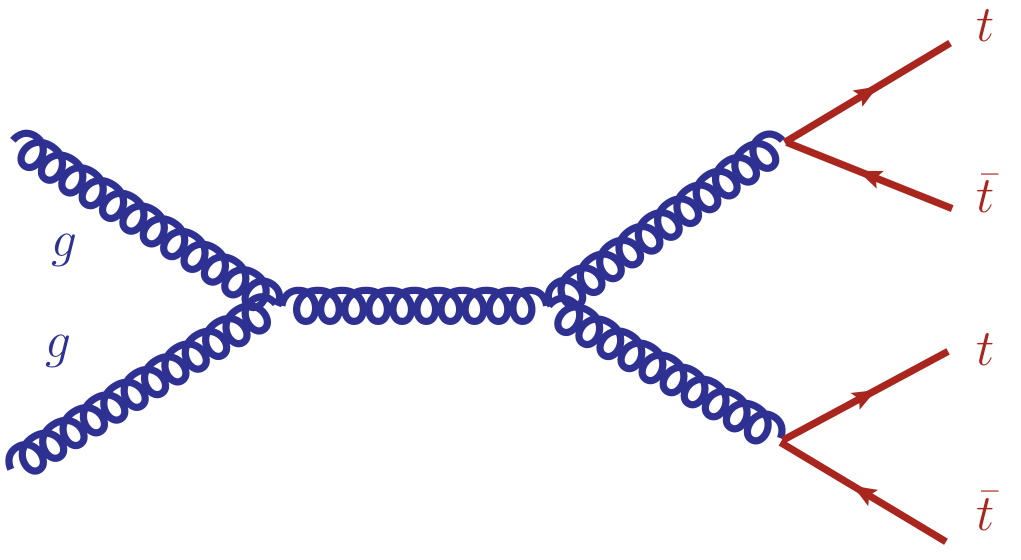}
\includegraphics[width=.22\textwidth,height=0.18\textwidth]{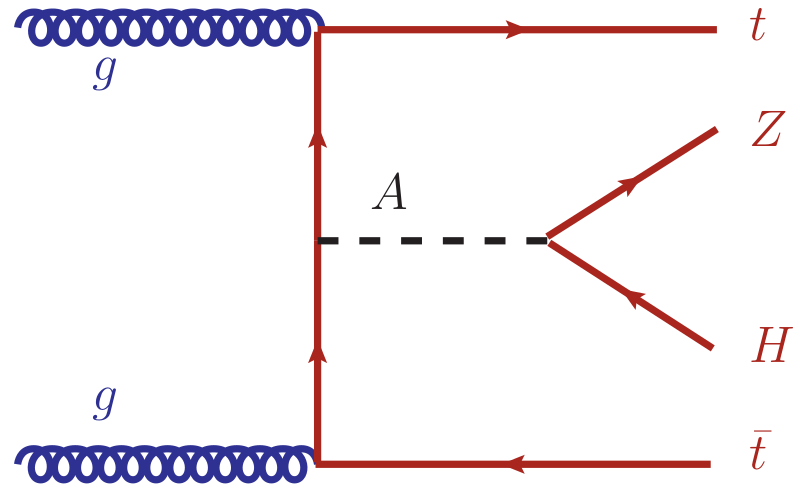}
\caption{\label{label1}The Feynman diagram for the four top quarks production at the leading order in QCD.}
\end{figure}

The large coupling of the top quark with the SM Higgs boson exposes it to multiple new particles as well as new interactions predicted in various physics in BSM. Thus, studying rare processes involving the top quarks are of high interest. The four top quark production has been observed recently by both ATLAS~\cite{ATLAS:2020hpj} and CMS~\cite{CMS:2019rvj} collaborations and it is one of the rare processes predicted by the SM. In this study we require two same-sign leptons (2LSS) or events with at least three leptons (3L) to be present in the final state. Although the branching fractions for these channels are relatively small, it delivers promising results owing to its clean nature and small background contribution from the SM. The addition of a Higgs-doublet to the SM resulted in the scalar spectrum being populated with two CP-even ($h,H$), one CP-odd ($A$) and charged scalar bosons ($H^{\pm}$), thus leaving room to study the characteristics of the scalar spectrum. 
Our interest is on investigating the CP-odd scalar in the 2HDM+$S$ model, by studying the production of $A$ in association with two top quarks and its decay into $A \rightarrow t\overline{t}$ channels. The relevant Feynman diagrams are shown in Fig.~\ref{label1}.


\section{The Model}
In order to understand the results reported by ATLAS~\cite{ATLAS:2020hpj}, we have considered a 2HDM extended with a real singlet scalar~\cite{vonBuddenbrock:2016rmr,vonBuddenbrock:2018xar}, $\Phi_S$, we kept the notation same as in Ref.~\cite{vonBuddenbrock:2016rmr} and name this model as 2HDM+$S$. The potential of the model is given by:

\begin{equation}\label{model:potential}
\begin{split}
V(\Phi_1, \Phi_2, \Phi_S) & = m_{11}^2 |\Phi_1|^2 + m_{22}^2 |\Phi_2|^2 - m_{12}^2 (\Phi_1^{\dagger}\Phi_2 + h.c.) \\
& + \frac{\lambda_1}{2} (\Phi_1^{\dagger}\Phi_1)^2 + \frac{\lambda_2}{2} (\Phi_2^{\dagger}\Phi_2)^2 + \lambda_3 (\Phi_1^{\dagger}\Phi_1) (\Phi_2^{\dagger}\Phi_2) \\
& + \lambda_4 (\Phi_1^{\dagger}\Phi_2) (\Phi_2^{\dagger}\Phi_1) + \frac{\lambda_5}{2} [(\Phi_1^{\dagger}\Phi_2)^2 + h.c.]\\
& + \frac{1}{2} m_S^2 \Phi_S^2 + \frac{\lambda_6}{8} \Phi_S^4 + \frac{\lambda_7}{2} (\Phi_1^{\dagger}\Phi_1) \Phi_S^2 + \frac{\lambda_8}{2} (\Phi_2^{\dagger}\Phi_2) \Phi_S^2.
\end{split}
\end{equation}

Here the $\Phi_1$ and $\Phi_2$ are the $SU(2)_L$ Higgs doublets and $\Phi_S$ is the singlet field. The first three lines correspond to the potential for the 2HDM and the last line is the contribution from the singlet field. In order to avoid the tree-level FCNC's, all the quarks of a given charge must couple to a single Higgs doublet, which can be done by imposing a $Z_2$ symmetry and it is softly broken by the $m_{12}^2$ term. Moreover, if we extend the $Z_2$ symmetry to the Yukawa sector it guarantees the absence of FCNC at tree level. Here, we consider a case where the real singlet field acquires a vacuum expectation value ($vev$) with a $Z_2$ symmetry. In other words, if this symmetry is respected then the singlet scalar becomes a viable dark matter candidate.
In this study we set $m_{12}^2 \ne 0$ in the 2HDM+$S$  potential, Eqn.~\ref{model:potential}, which corresponds to a soft breaking of the $Z_2$ symmetry. Since we do not consider explicit CP violation, we assume $\lambda_i$ to be real. Minimising the potential with the three Higgs fields and assuming the $vevs$ of the fields $\Phi_1 \rightarrow v_1/\sqrt{2}$, $\Phi_2 \rightarrow v_2/\sqrt{2}$ and $\Phi_S \rightarrow v_S$, the three minimisation conditions are:

\begin{equation}\label{minimisation_cond}
\frac{\partial V}{\partial [v_1, v_2, v_S]} = 0.
\end{equation}
With this the conditions on the fields are:
\begin{eqnarray}
m_{11}^2 &= -\frac{1}{2}(v_1^2\lambda_1 + v_2^2\lambda_{345} + v_S^2\lambda_7) + \frac{v_2}{v_1}m_{12}^2, \\
m_{22}^2 &= -\frac{1}{2}(v_2^2\lambda_2 + v_1^2\lambda_{345} + v_S^2\lambda_8) + \frac{v_2}{v_1}m_{12}^2, \\
m_{S}^2 &= -\frac{1}{2}(v_1^2\lambda_7 + v_2^2\lambda_8 + v_S^2\lambda_6),
\end{eqnarray}
where $\lambda_{345} = \lambda_3 + \lambda_4 + \lambda_5$.

\section{Deep Neural Networks}
\begin{figure}[t]
		\includegraphics[width=0.45\textwidth]{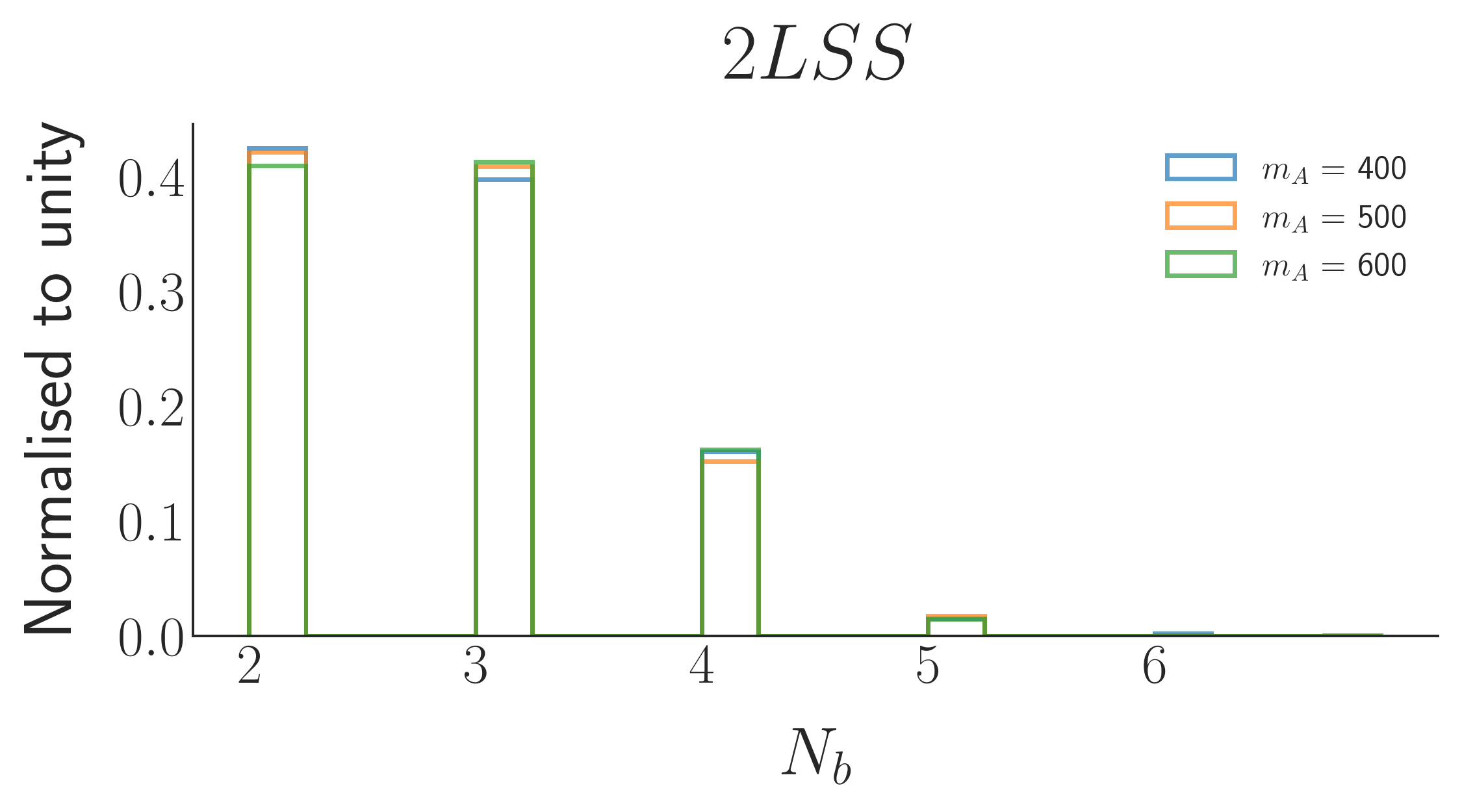}
		\includegraphics[width=0.45\textwidth]{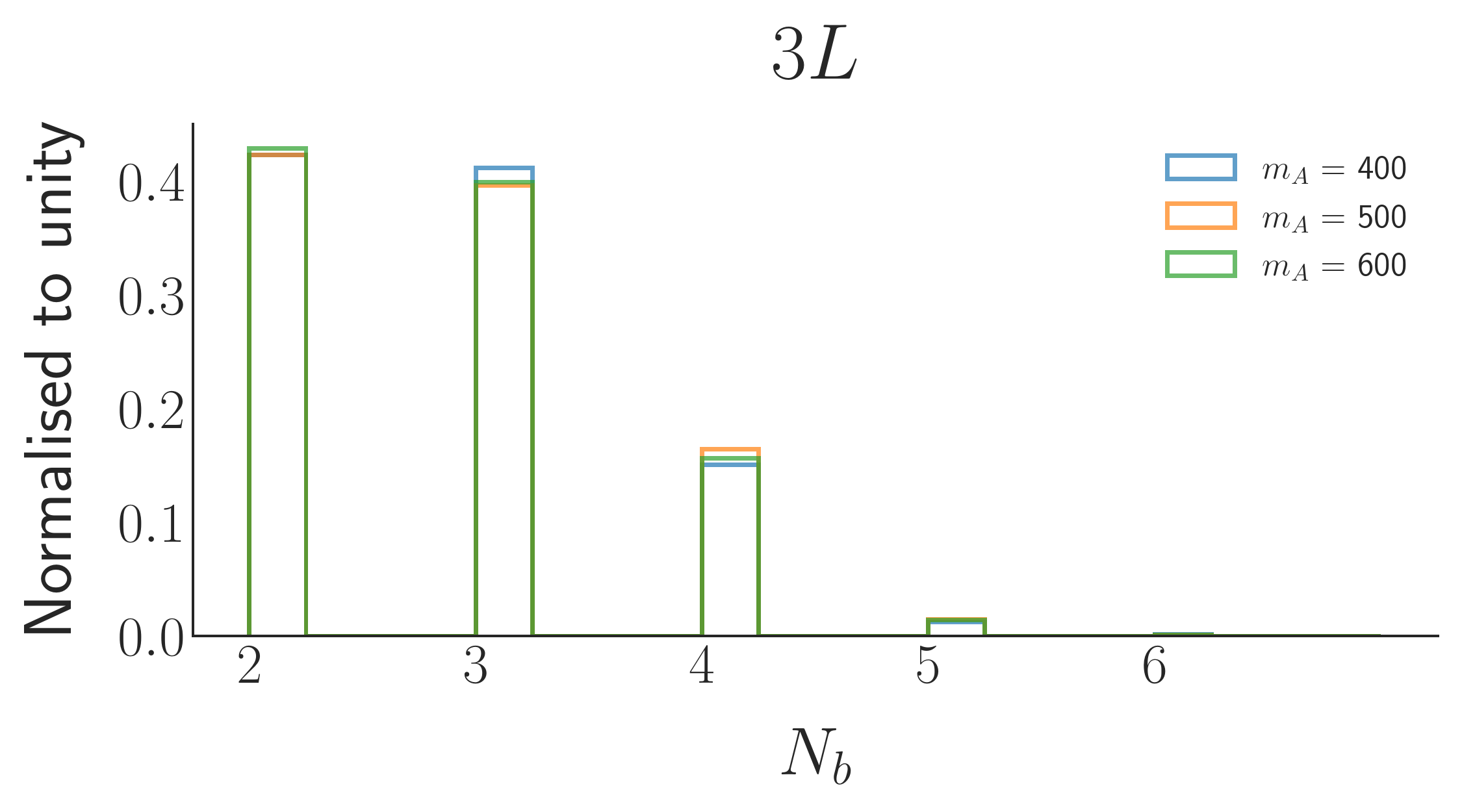}
\caption{\label{label2} The $b$-jet multiplicity in events from $pp\rightarrow t\overline{t}A\rightarrow t\overline{t}t\overline{t}$ production for three values of $m_A$ with the event selection applied in Ref.~\cite{ATLAS:2020hpj}. The graphs on the left (right) correspond to two same-sign (three) leptons.}
\end{figure}

Deep Neural Networks (DNN) are considered a powerful tool for large-scaled problems when using machine learning. They have been applied to a variety of problems, essentially classifications of various types and to Artificial Intelligence complex systems needed at CERN when searching for new Physics. Large data sets with various kinematics are produced by the multipurpose detectors at the LHC from the proton-proton collisions. To get information from the data, different mechanisms are used to process the multi-dimensional space.

DNNs make use of multiple processing layers to determine patterns within large data sets, through each layer learning from the input data and subsequently passing on the information to the next layer. The data flow in DNNs is in one direction, from the input layer to the output layer with the connections between the layers being fed forward, while the output results are obtained using back propagation.

In our study we have engineered three separate DNN models that will be used to separate the SM production of four top quarks from the BSM production, as we have evaluated the heavy pseudo scalar $A$ with three different masses, $m_A$ = 400, 500, 600\,GeV. The best working hyper parameters for the DNNs were used to build DNNs that provided efficient results for the classification task. 

\section{Results} 

\begin{figure}[t]
\centering
        \includegraphics[width=.30\textwidth]{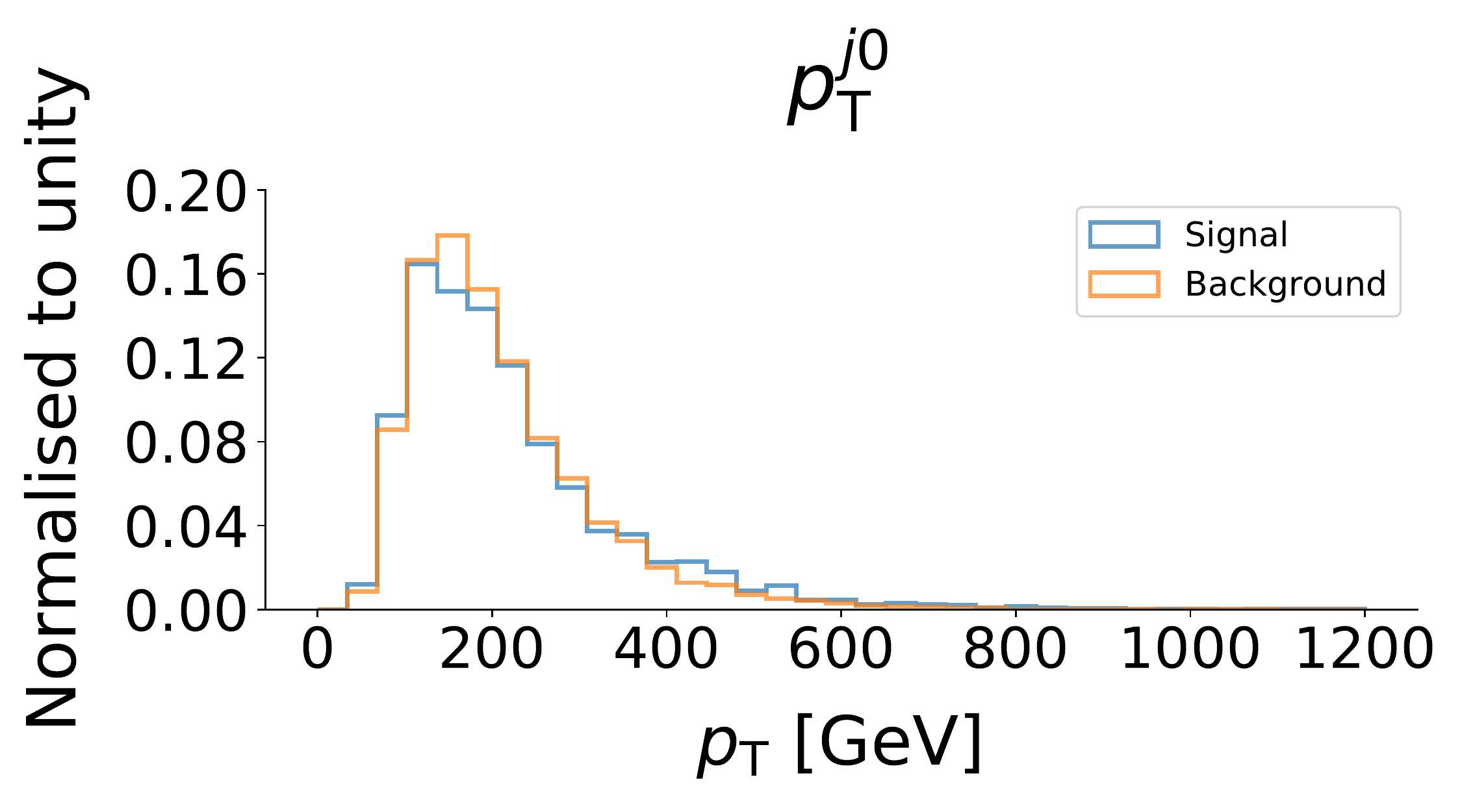}
        \includegraphics[width=.30\textwidth]{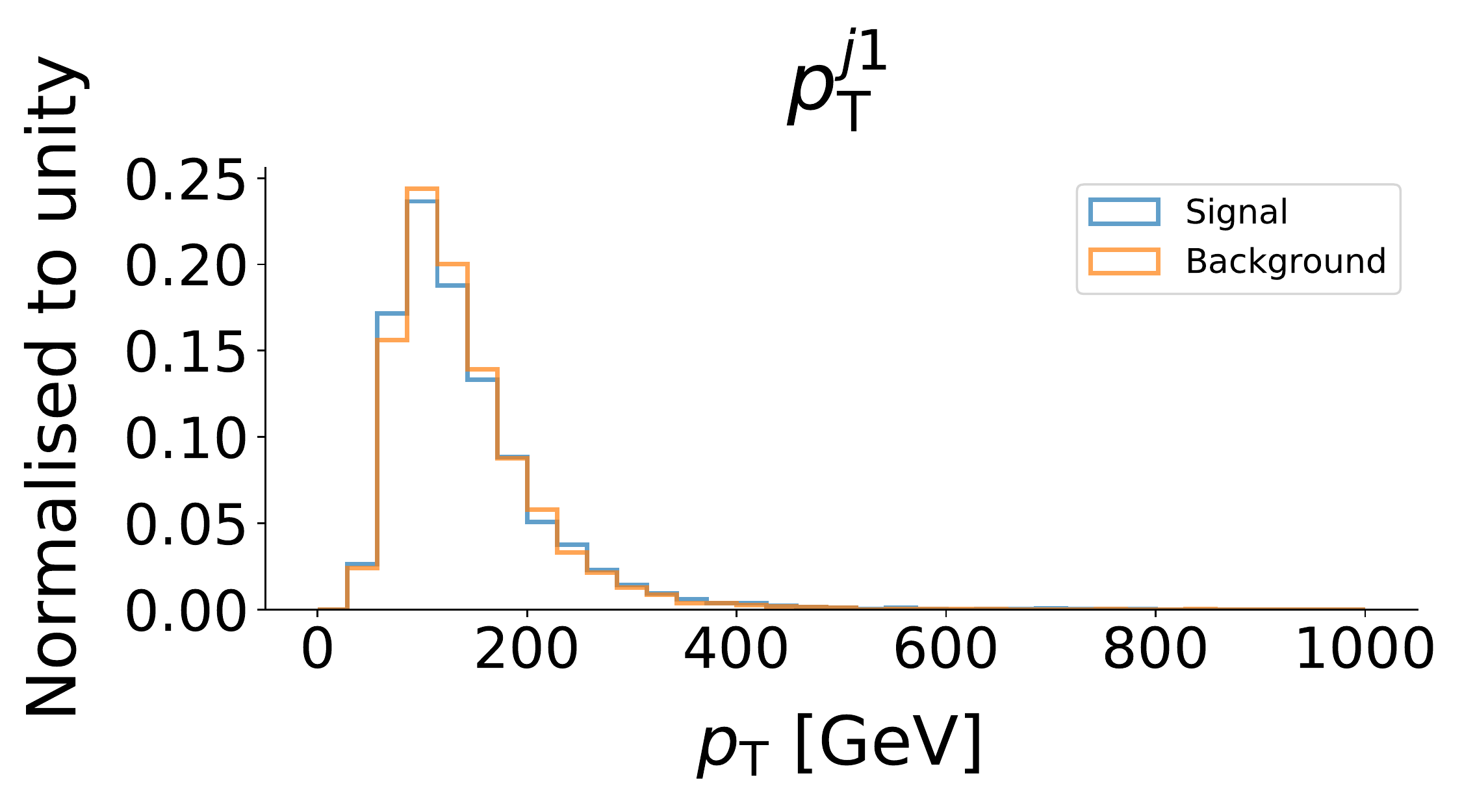}
        \includegraphics[width=.30\textwidth]{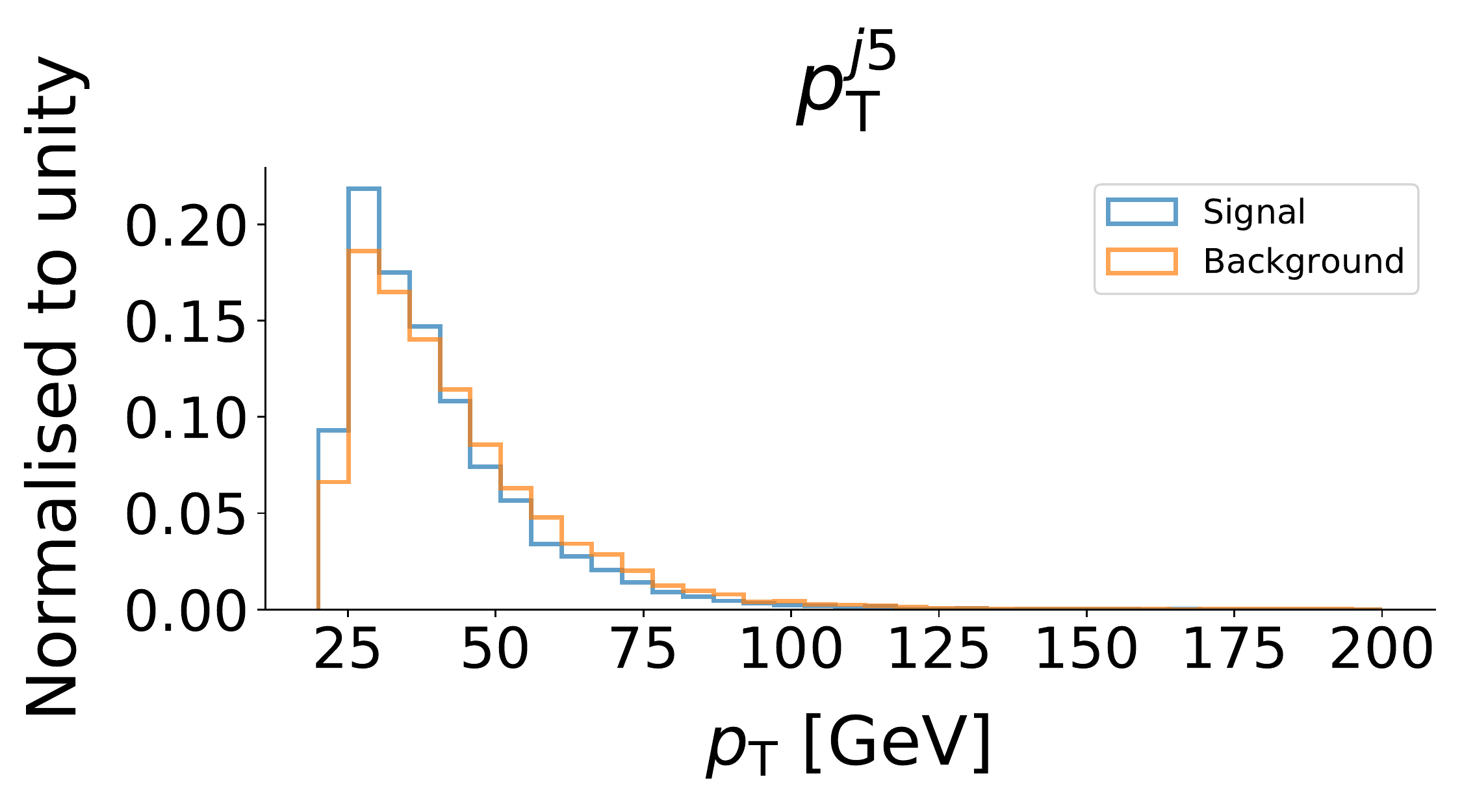}
        \includegraphics[width=.30\textwidth]{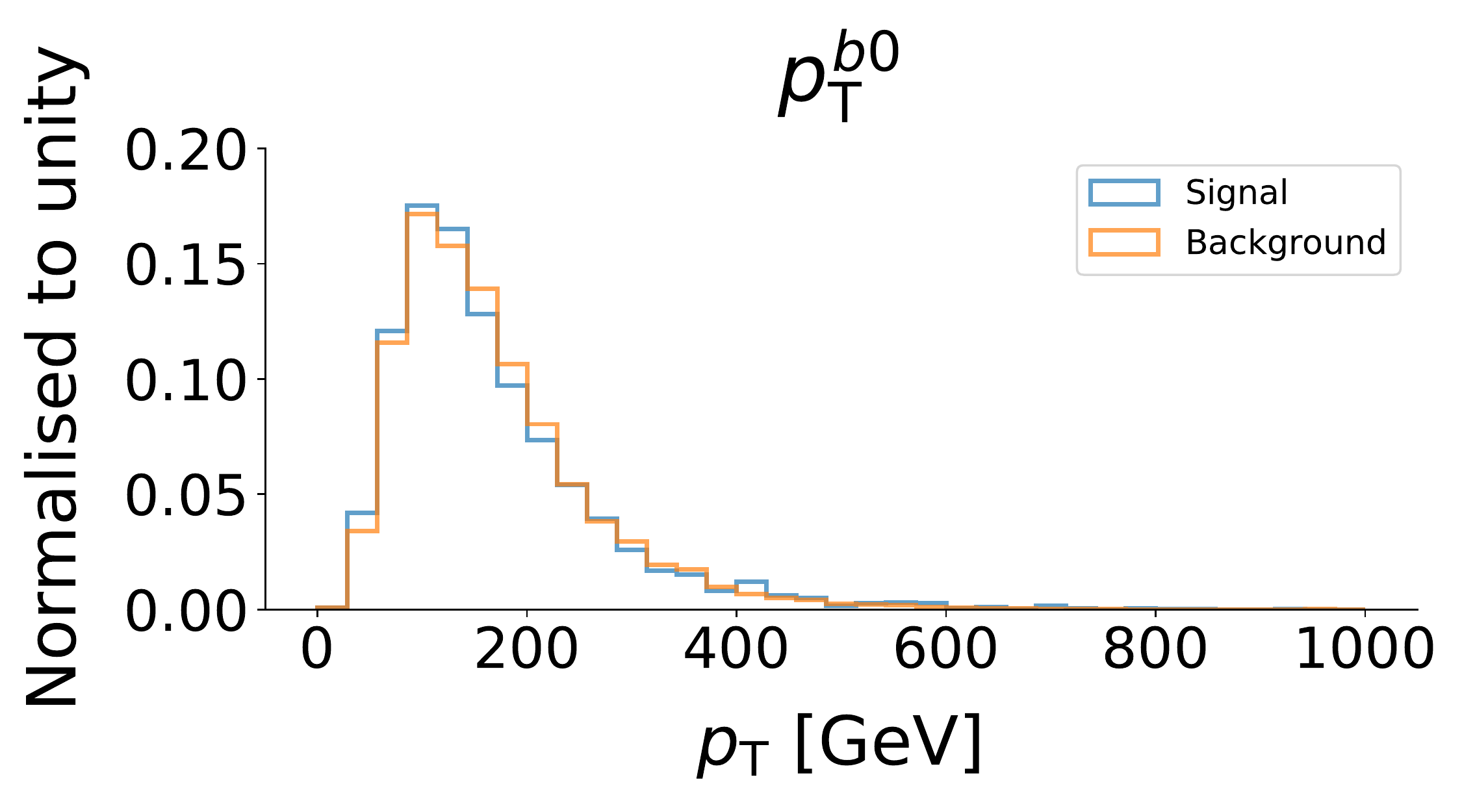}
        \includegraphics[width=.30\textwidth]{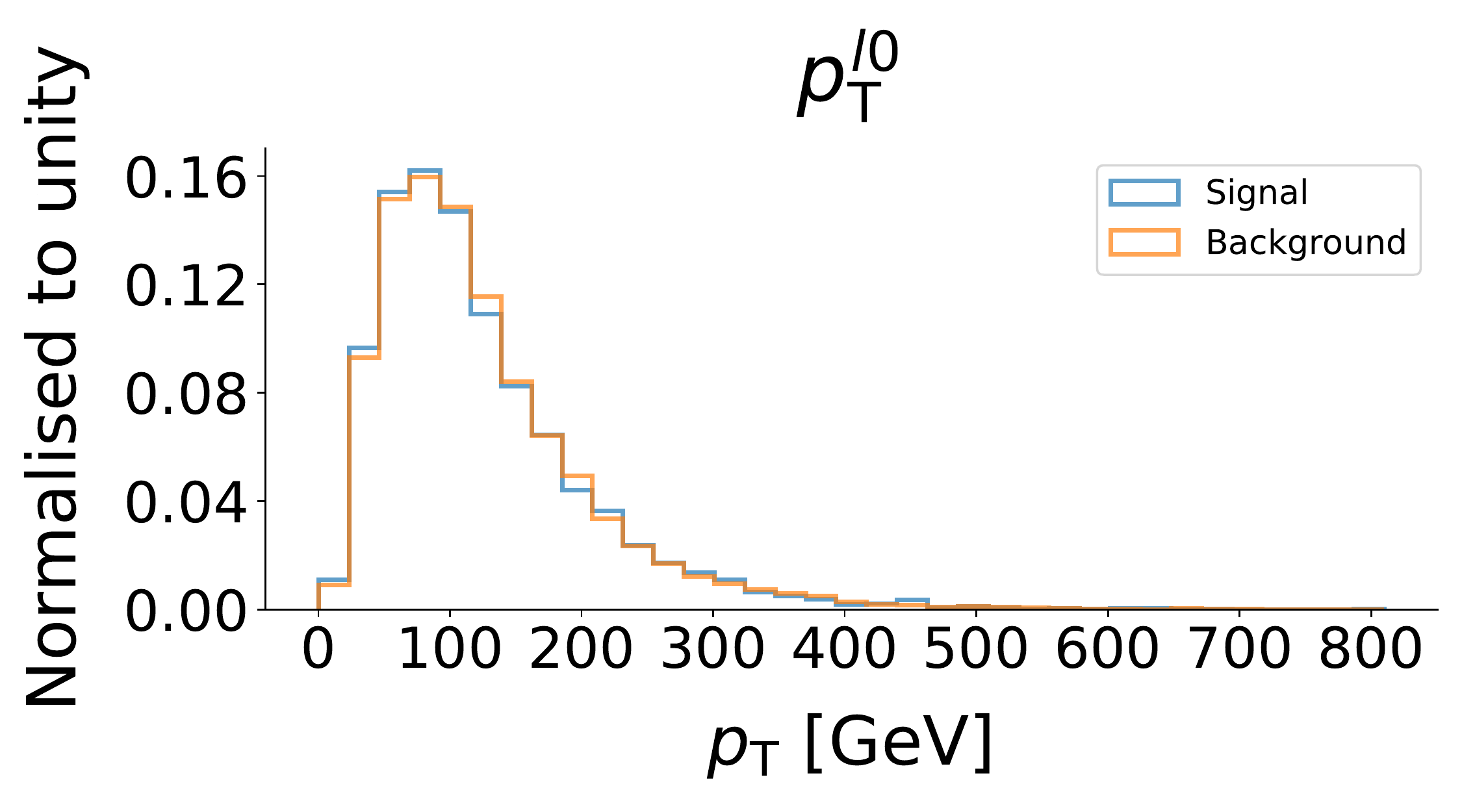}
        \includegraphics[width=.30\textwidth]{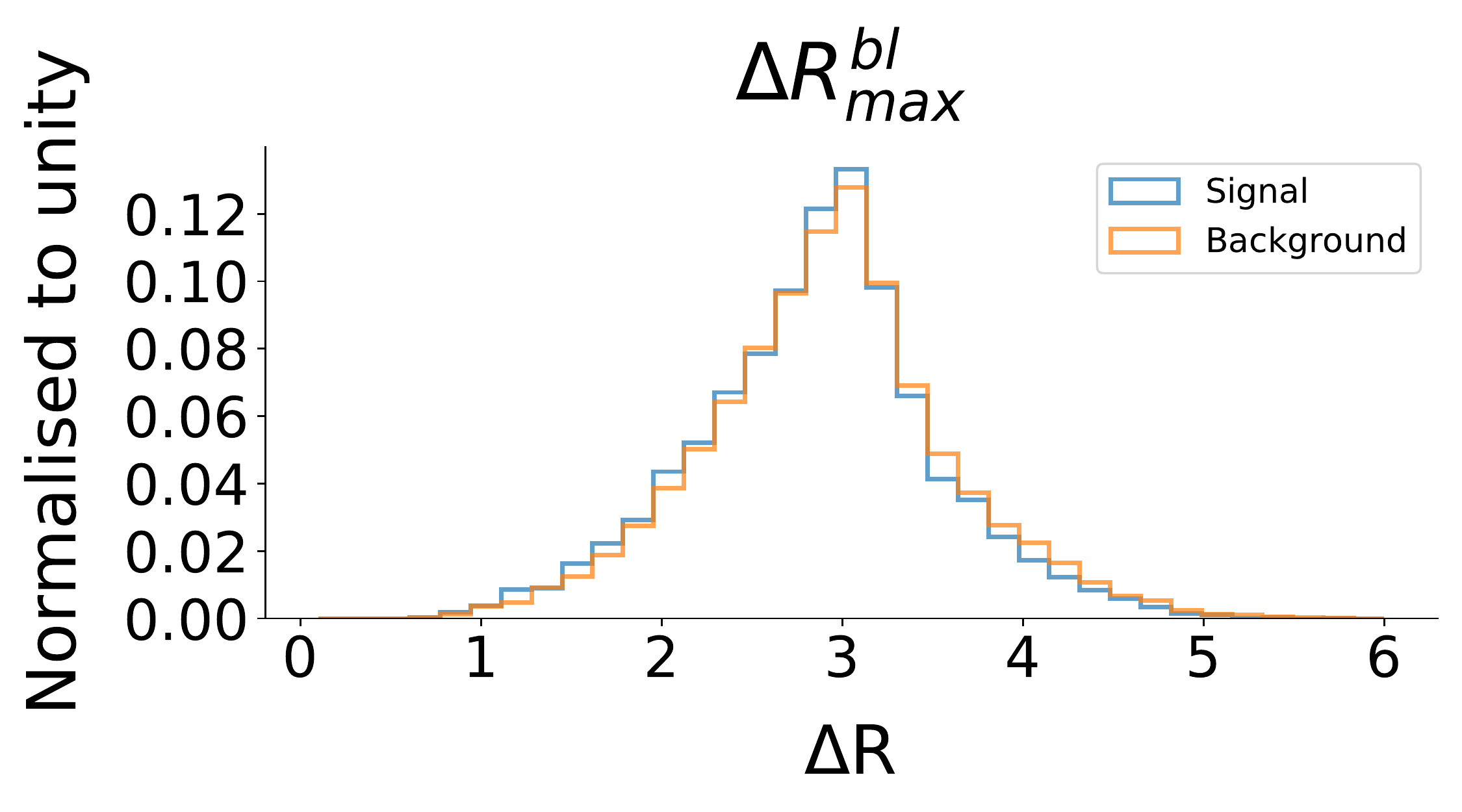}
        \includegraphics[width=.30\textwidth]{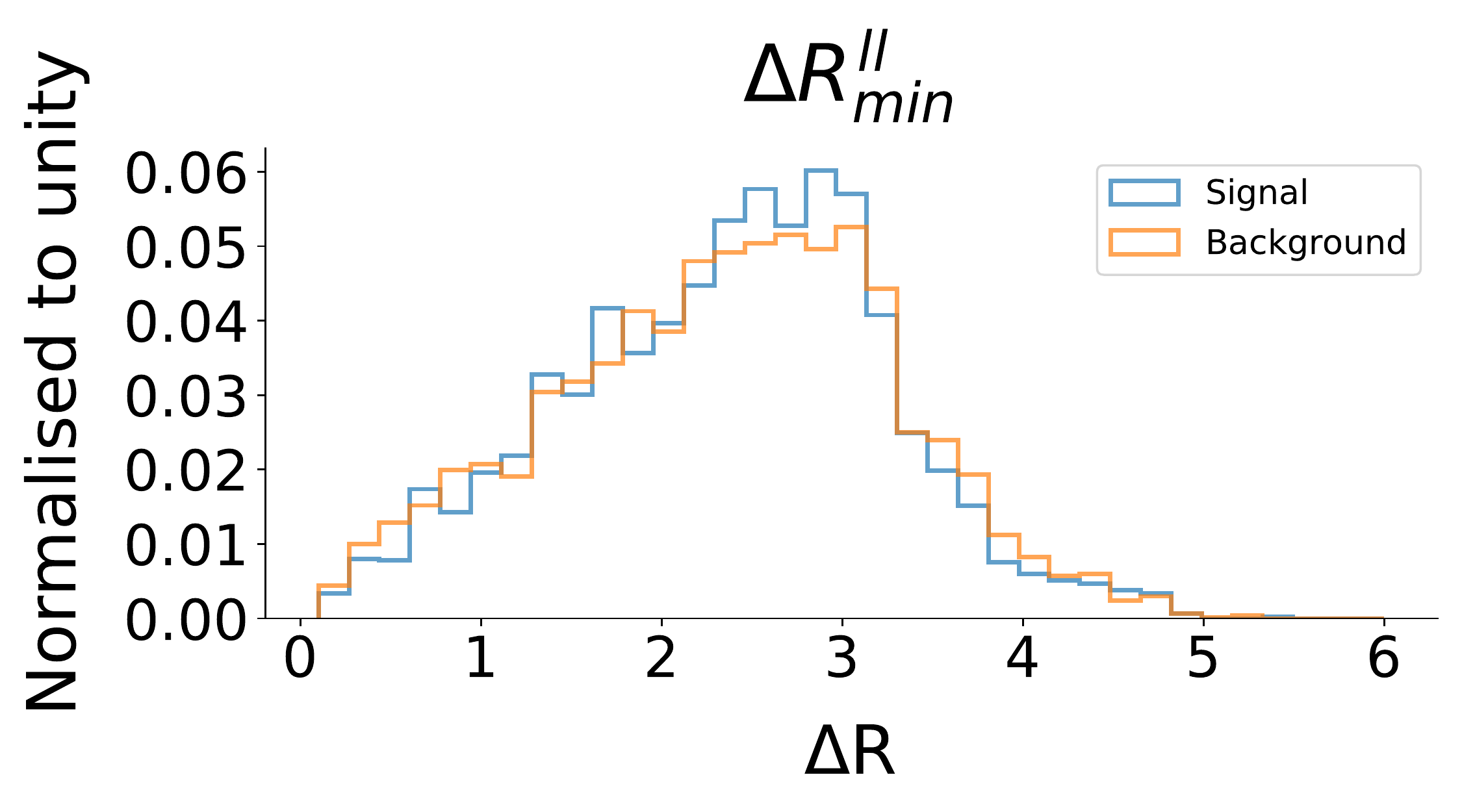}
        \includegraphics[width=.30\textwidth]{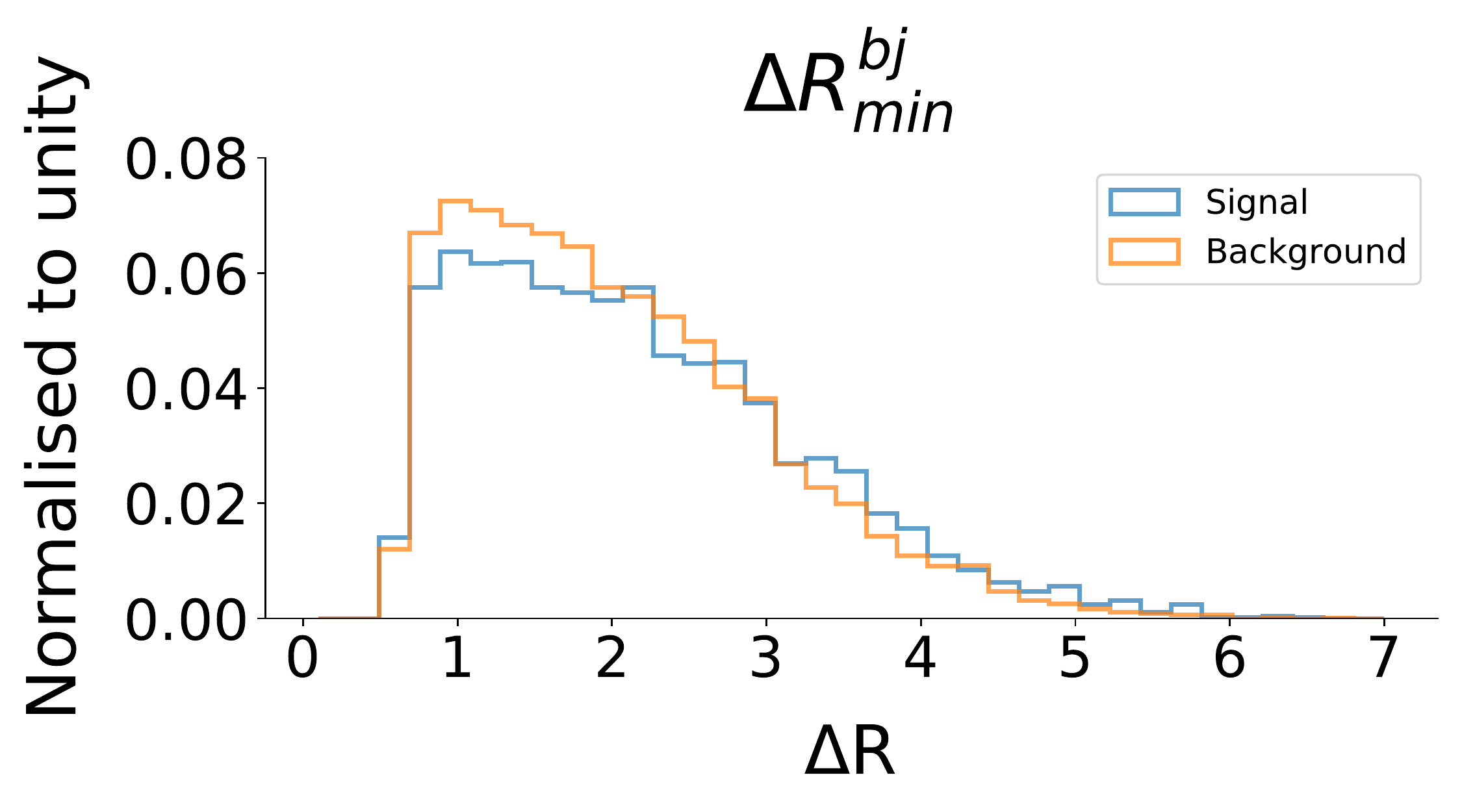}
        \includegraphics[width=.30\textwidth]{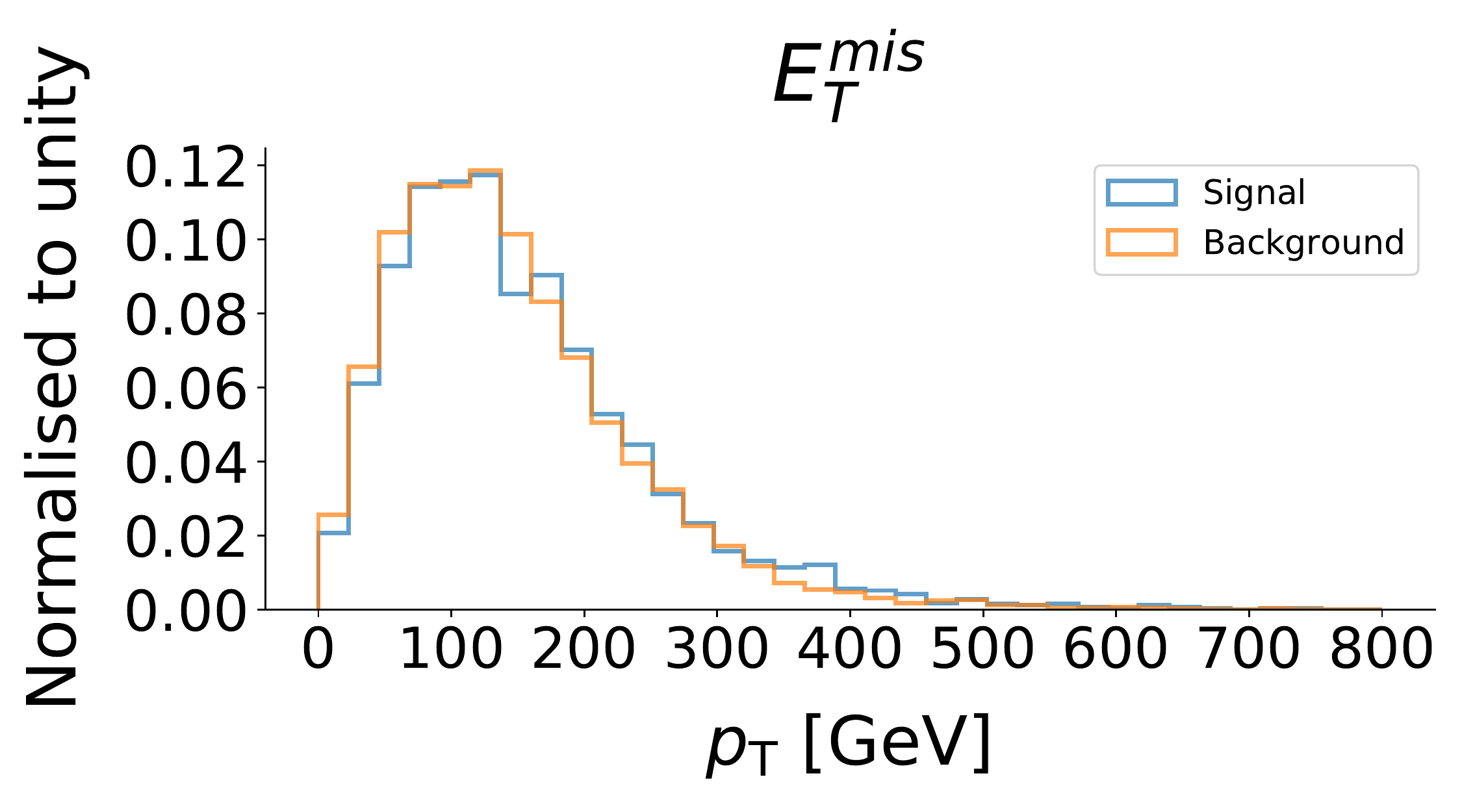}
        \includegraphics[width=.30\textwidth]{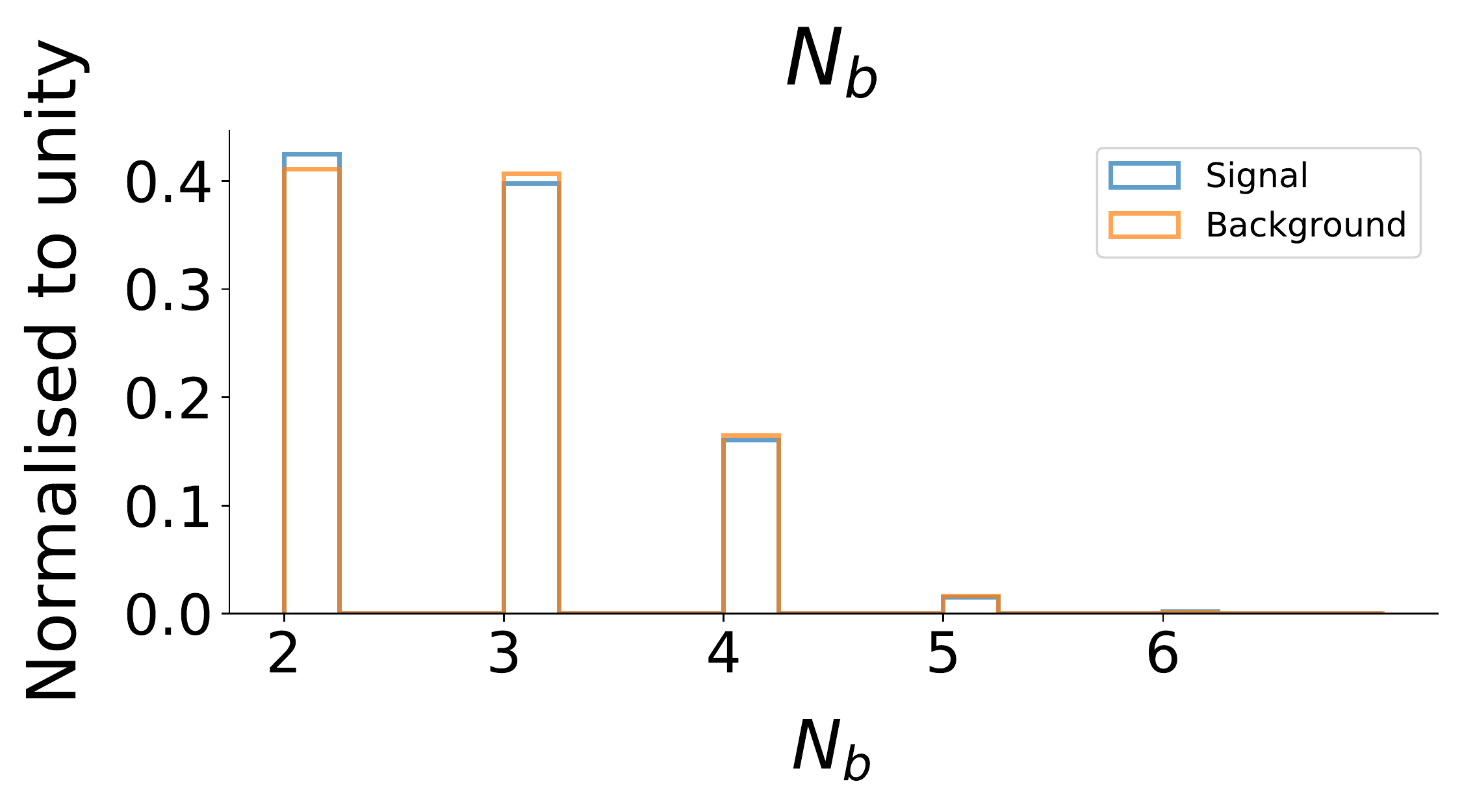}
        \includegraphics[width=.30\textwidth]{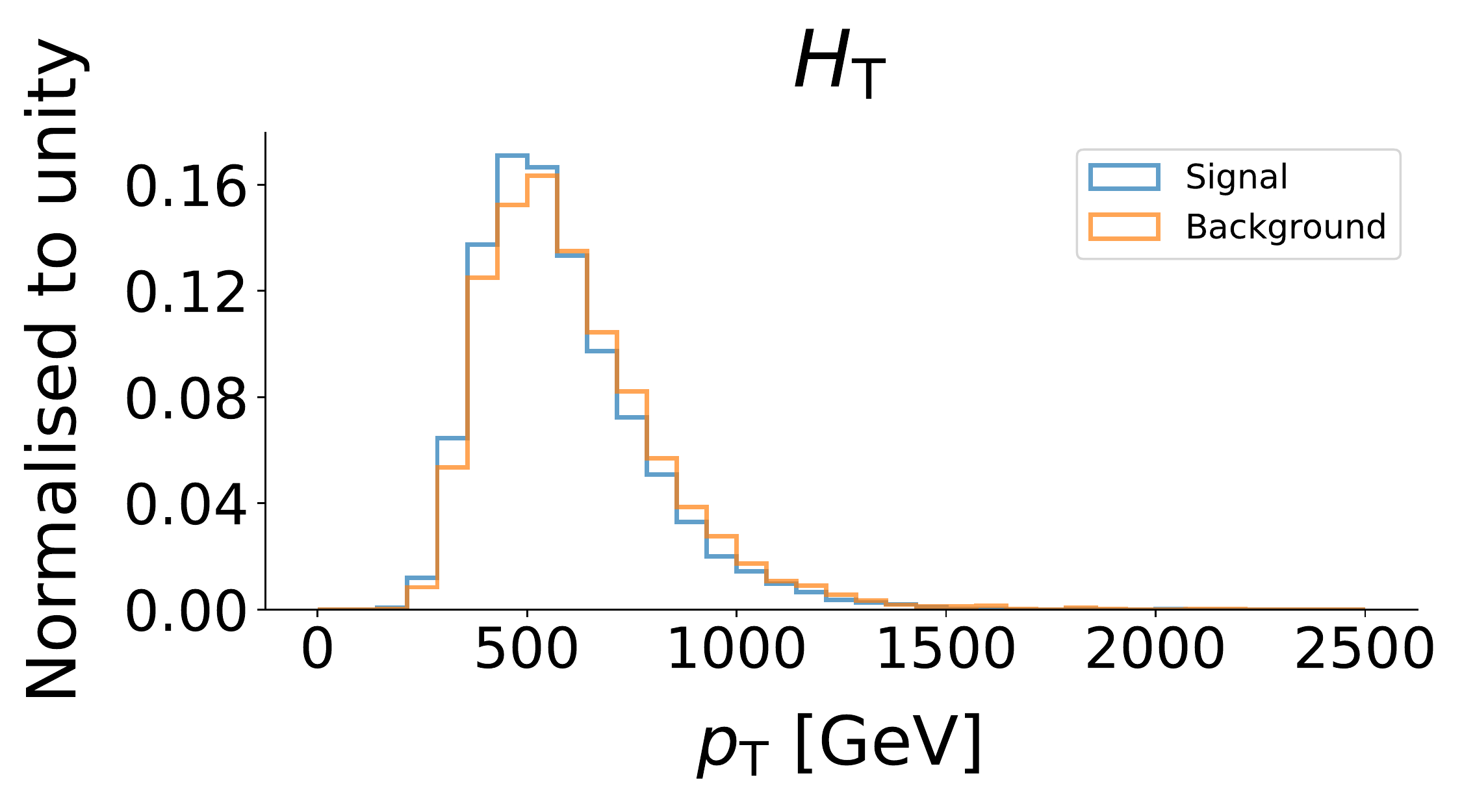}
        \includegraphics[width=.30\textwidth]{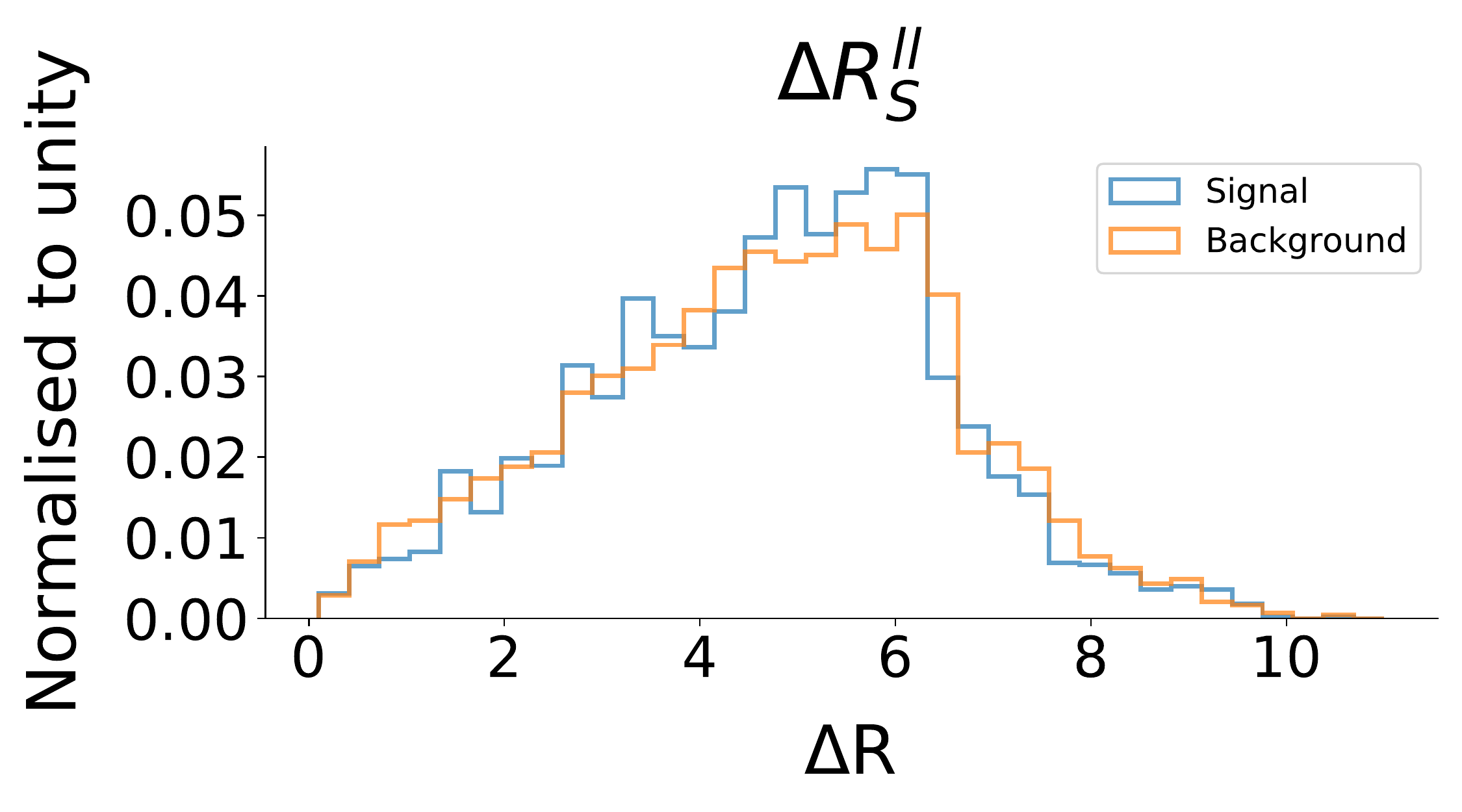}
\caption{\label{label3}The distributions of the input variables used to train the DNN for two same-sign leptons. The BSM signal corresponds to $m_A=400$\,GeV.} 
\end{figure}

 We perform a Monte-Carlo simulation of $pp$ collisions at the LHC. The events corresponding to the signal and SM backgrounds are generated using {\tt Madgraph5}~\cite{Alwall:2014hca} with the {\tt NNPDF3.0} parton distribution functions~\cite{NNPDF:2014otw}. The UFO model files required for the Madgraph analysis have been obtained from {\tt FeynRules}~\cite{Alloul:2013bka} after a proper implementation of the model. Following this parton\-level analysis, the parton showering and hadronization are performed using {\tt Pythia}~\cite{Sjostrand:2006za}. We use {\tt Delphes}(v3)~\cite{deFavereau:2013fsa} for the corresponding detector level simulation after the showering/hadronization.

The event selection documented in Ref.~\cite{ATLAS:2020hpj} is used here
The DNN used in this study is a binary classification algorithm which categories between 0 and 1, with 0 being associated with the SM production of four top quarks while 1 is associated with BSM production. The signature that will be used to explain the results reported by ATLAS in Ref.~\cite{ATLAS:2020hpj} with $b$-jets follows the production mechanism $pp\rightarrow t\overline{t}A\rightarrow t\overline{t}t\overline{t}$.  This gives rise to an excess of multi-lepton final states associated with $b$-tagged jets in the two channels of interest. In order to illustrate the excess of $b$-tagged jets in the final state of the signature that we are studying, distributions of the $b$-tagged jets for the two channels are shown in Fig.~\ref{label2}.

The discriminating features used to train the DNN in order to perform a classification between the SM and BSM four top quark productions are displayed in Fig.~\ref{label3}. This  includes the total number of $b$-tagged jets, $N_b$, the leading lepton transverse momentum, $p_{T}^{l0}$, the missing transverse energy, $E_{T}^{mis}$, the leading jet transverse momentum, $p_{T}^{j0}$, the second leading jet transverse momentum, $p_{T}^{j1}$, the sixth leading jet transverse momentum, $p_{T}^{j5}$, the leading $b$-tagged jet transverse momentum, $p_{T}^{b0}$, the minimum distance defined as $\Delta R = \sqrt{(\Delta \eta)^2 + (\Delta \phi)^2}$ between two leptons out of all possible pairs, $\Delta R_{min}^{\ell \ell}$, the scalar sum of transverse momenta over all leptons and jets excluding the leading jet, $H_T$, the sum of the distances between two leptons for all possible pairs, $\Delta R_{S}^{\ell \ell}$, the maximum distance between a $b$-tagged jet and a lepton among all possible pairs $\Delta R_{max}^{b \ell}$, and the minimum distance between a jet and a $b$-tagged jet among all possible pairs $\Delta R_{min}^{b j}$. From the distributions of these input variables we are able to see that there is not much discrimination between the background  and the signal.

\begin{figure}[t]
\centering
		\includegraphics[width=0.30\textwidth]{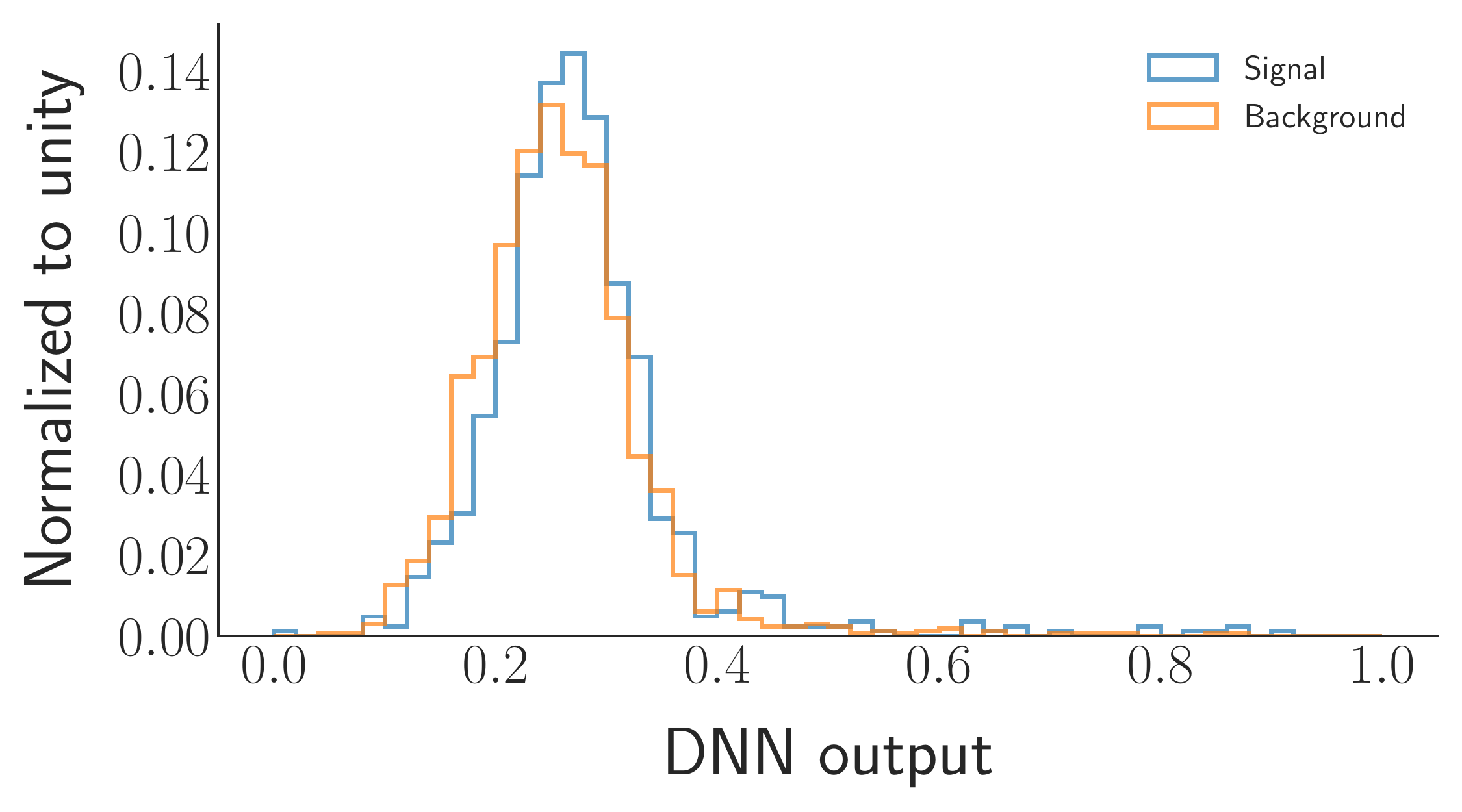}
		\includegraphics[width=0.30\textwidth]{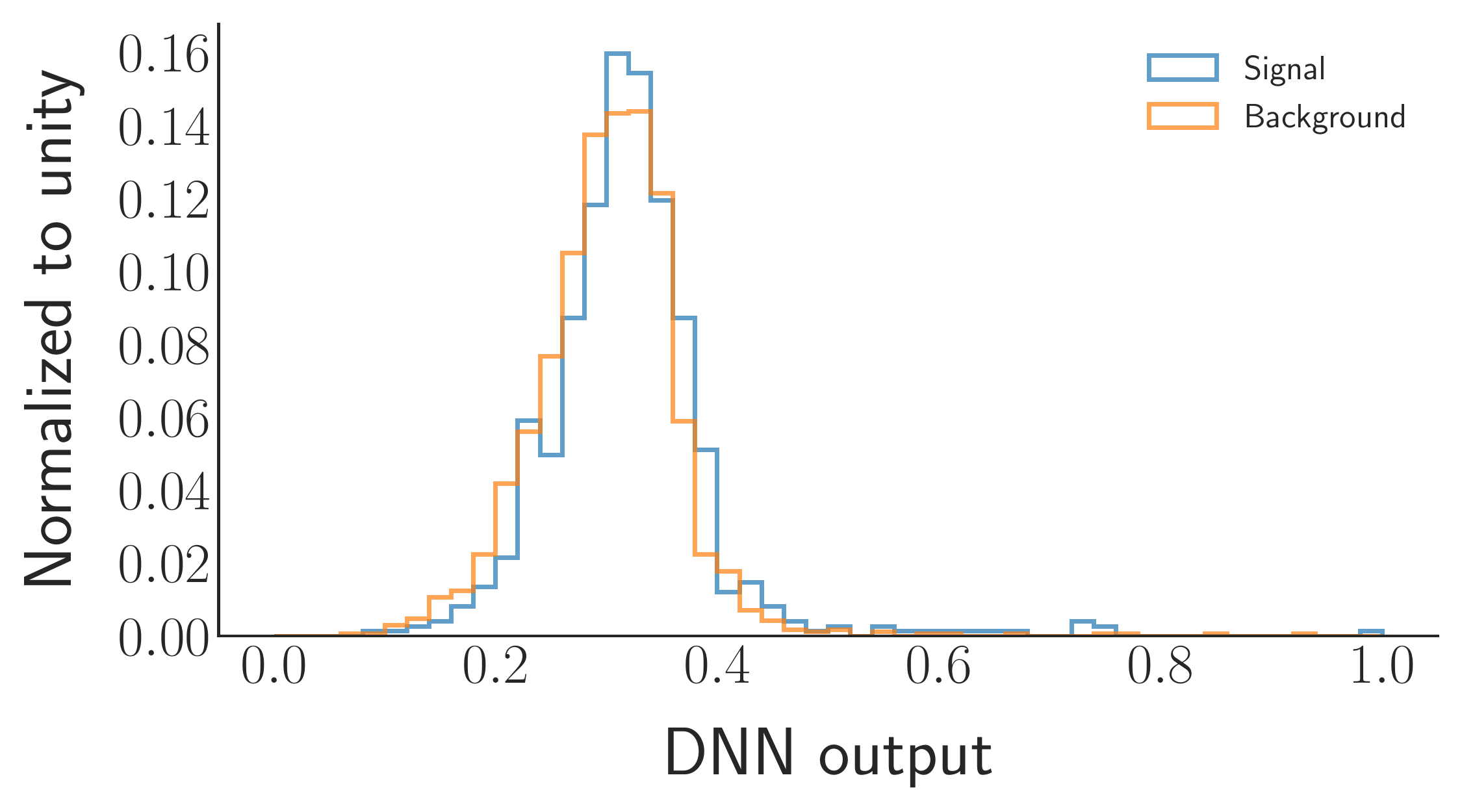}
		\includegraphics[width=0.30\textwidth]{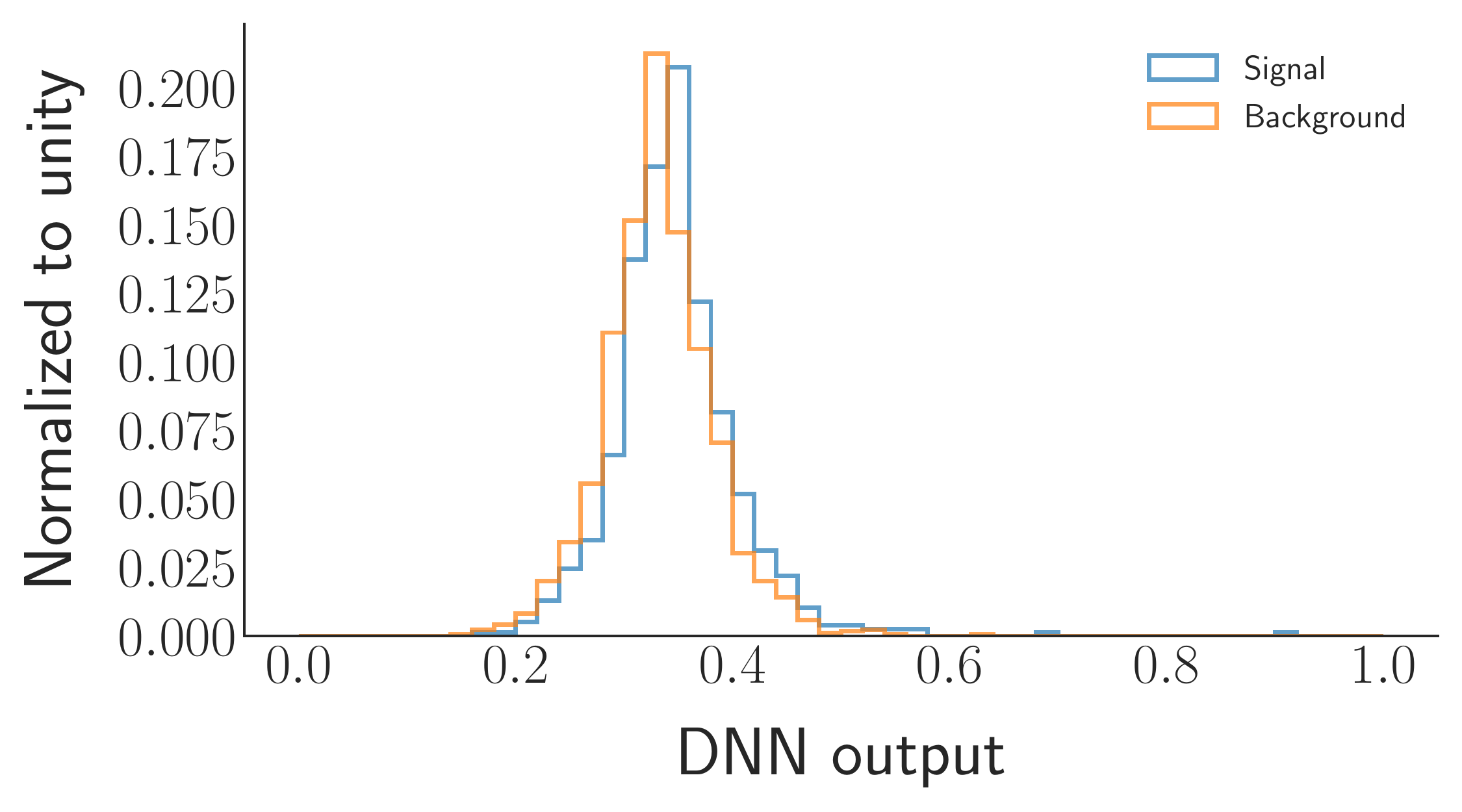}
		\includegraphics[width=0.30\textwidth]{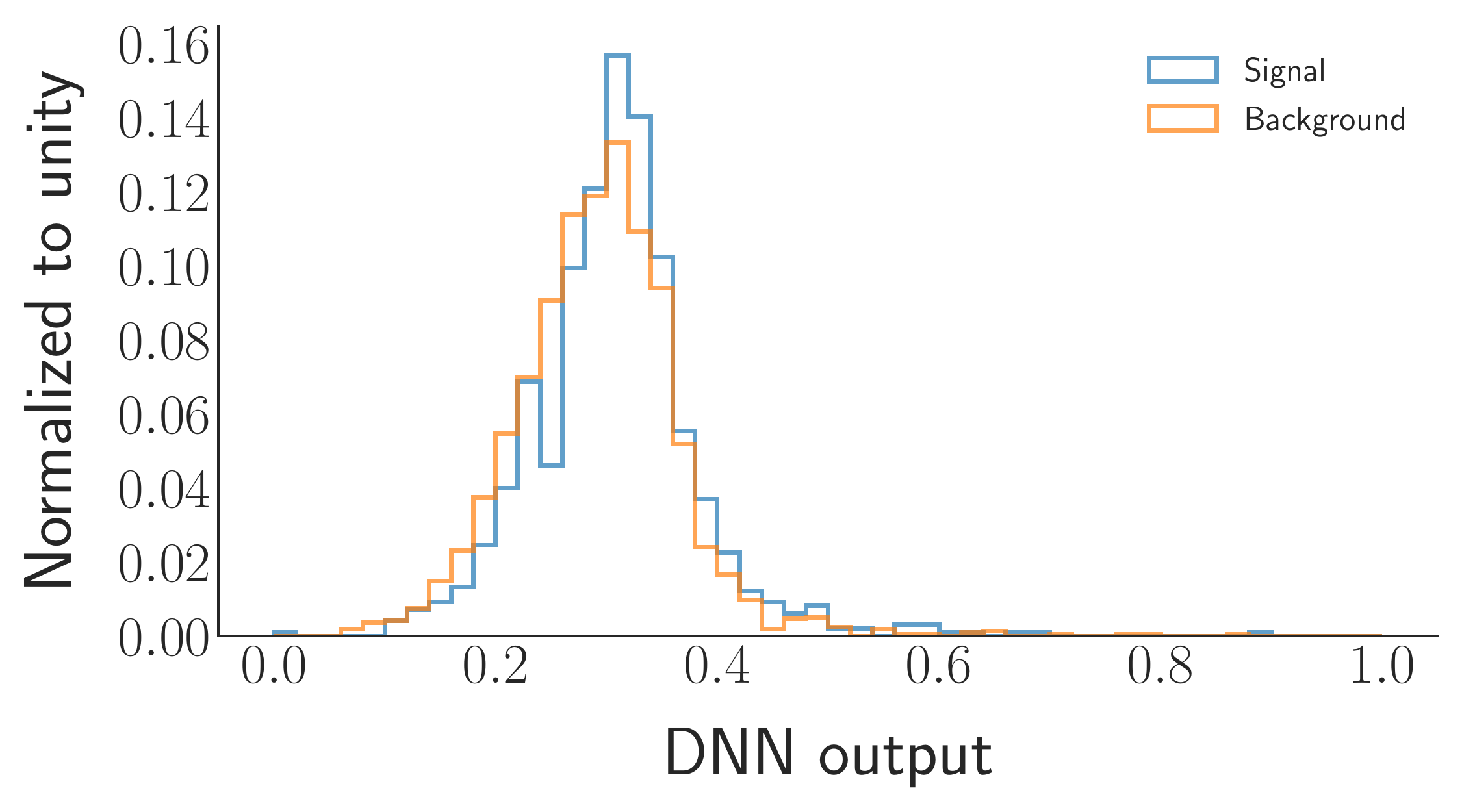}
		\includegraphics[width=0.30\textwidth]{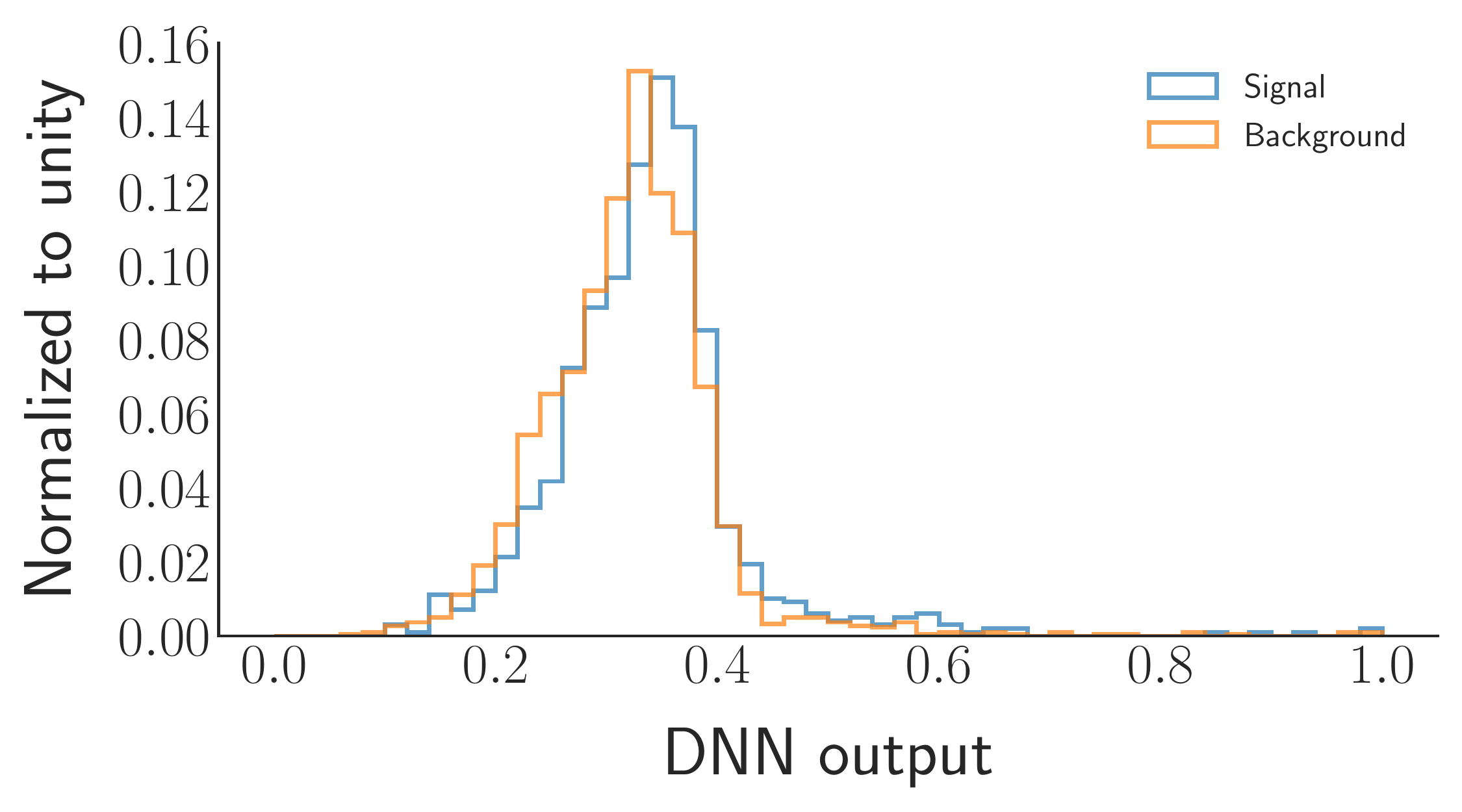}
		\includegraphics[width=0.30\textwidth]{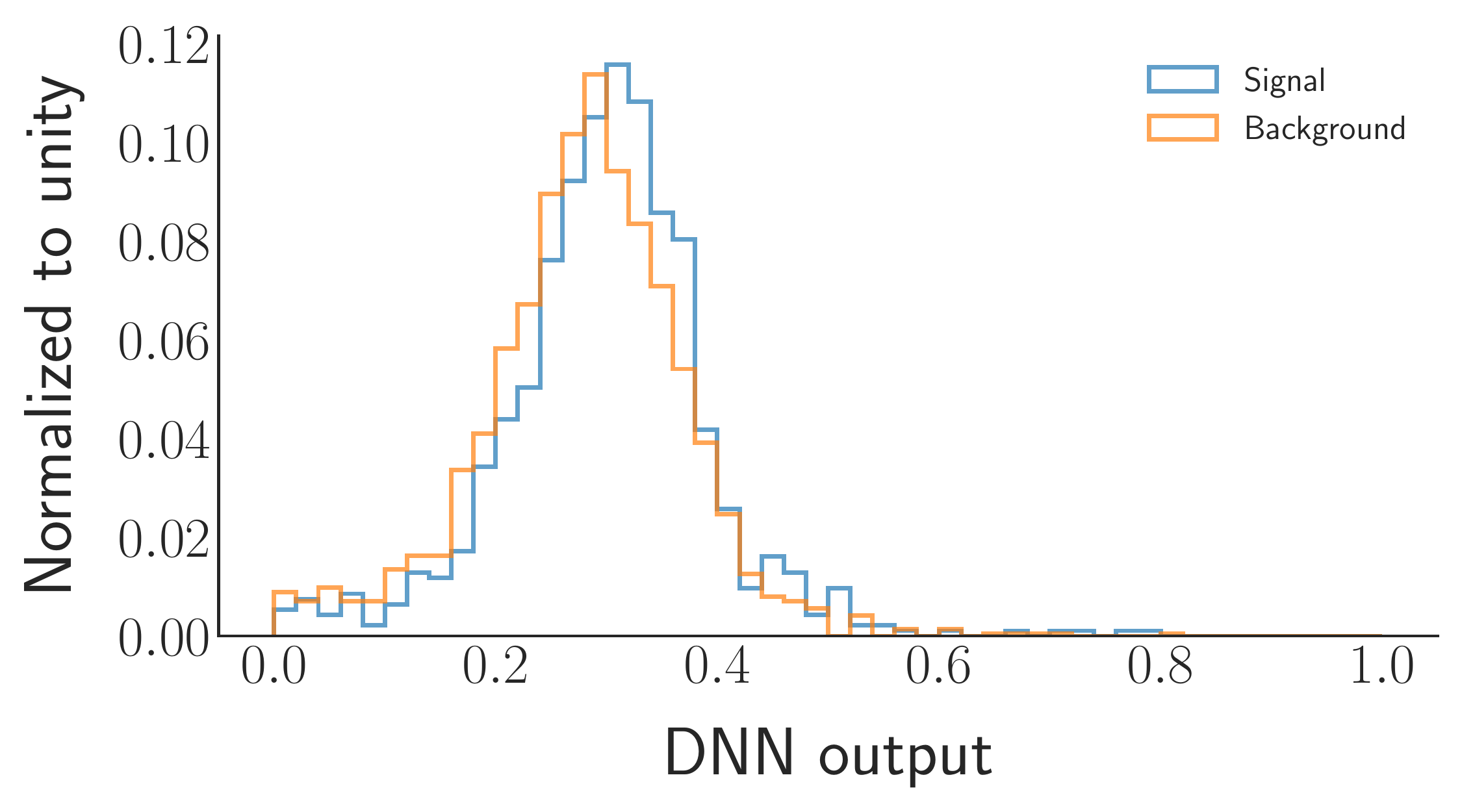}
\caption{\label{label4} The DNN output distributions for the three mass ranges for two channels of interest. The first row represents SS and the second row represents 3L. The first column is for $m_{A}$=400\,GeV, the second column is for $m_{A}$=500\,GeV and the third column is for $m_{A}$=600\,GeV.}
\end{figure}

\begin{figure}[t]
\centering
		\includegraphics[width=0.45\textwidth]{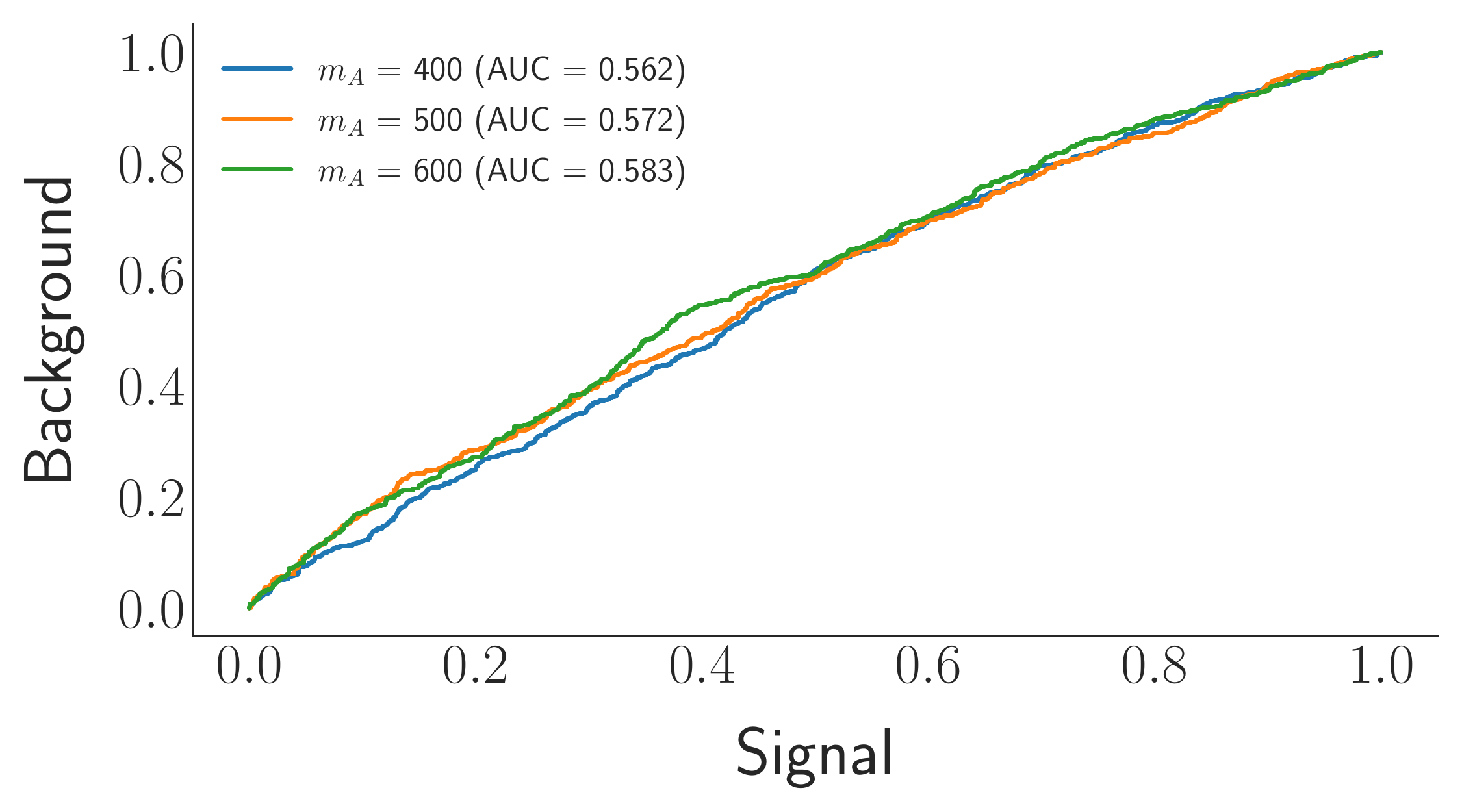}
		\includegraphics[width=0.45\textwidth]{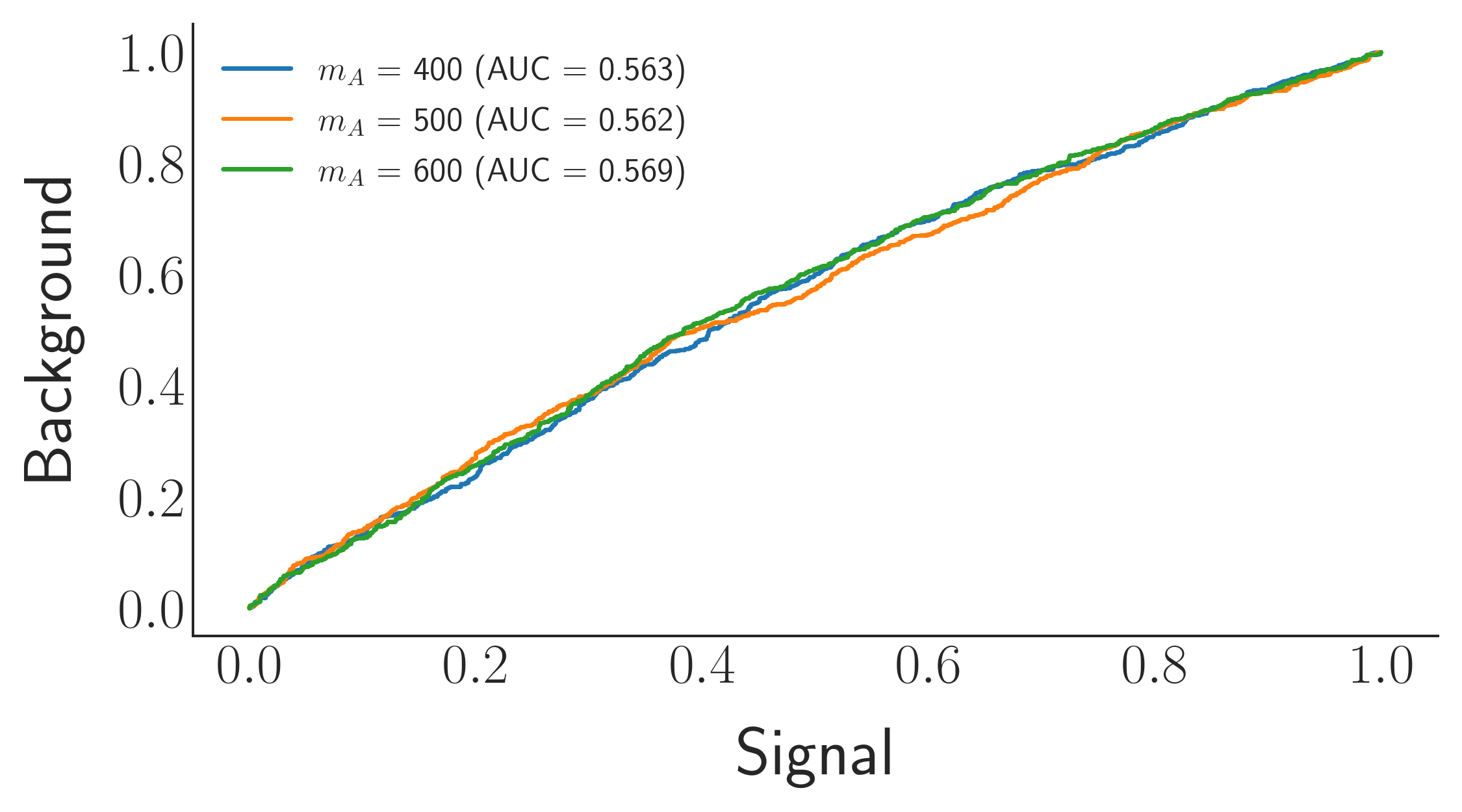}
\caption{\label{label5} The ROC curves obtained from the DNN models for signal and background. The graphs correspond to 2LSS (left) and 3L (right).}
\end{figure}

The DNN was set to the best hyper parameters and output distributions obtained from the model for testing data are shown in Fig.~\ref{label4}. From the output distributions of the DNN, we see that the SM four top quark and BSM four top quark productions are not far from each other, thus making it difficult to separate the two. This is further illustrated by the ROC (receiving operating characteristic) curves displayed in Fig.~\ref{label5} whose AUC (area under curve) are barely above 50\%. The ROC curves are obtained with the corresponding test data samples, which where not used in the training of the DNNs.

\section{Summary and Conclusion}

We have studied the four top quark production at the center of mass $\sqrt{s}=13$\,TeV at the  LHC  with two categories of multi-lepton channels: two same-sign leptons and three leptons. After using a number of kinematic variables, we notice that there is not much discrimination between the four top quark production in the SM and that of the BSM model used here ($pp\rightarrow t\overline{t}A\rightarrow t\overline{t}t\overline{t}$). A multivariate analysis is performed with a DNN using twelve  features, where no significant discrimination between the SM and BSM four top quark signals. This is illustrated by the AUCs of the ROC curves being marginally higher than 50$\%$.
Hence, we are predicting that the BSM signal should be seen as an elevation of the measured four top cross-section.

%
%
\section*{References}
\bibliographystyle{iopart-num}
\bibliography{Reference}

\end{document}